\documentclass[useAMS,usenatbib]{mn2e}
\usepackage{graphics}
\usepackage{epsfig}
\usepackage{graphicx}

\newcommand{\apj}{ApJ}

\newcommand{\mnras}{MNRAS}

\newcommand{\aap}{A\&A}

\newcommand{\apjl}{ApJL}

\newcommand{\nat}{Nature}

\newcommand{\mbh}{M_{\rm bh}}
\def\ltsima{$\; \buildrel < \over \sim \;$}
\def\simlt{\lower.5ex\hbox{\ltsima}}
\def\gtsima{$\; \buildrel > \over \sim \;$}
\def\simgt{\lower.5ex\hbox{\gtsima}}

\def\msun{{\,{\rm M}_\odot}}

\def\del#1{{}}

\title[Quasar feedback accelerate star formation]{Quasar feedback: accelerated star formation
  and chaotic accretion}

\author[S. Nayakshin, K. Zubovas]{Sergei Nayakshin and Kastytis Zubovas\\ 
Department of Physics \& Astronomy, University of Leicester, Leicester, LE1 7RH, UK\\
{E-mail:~} {\rm Sergei.Nayakshin@astro.le.ac.uk}}

\begin{document}

\date{Received}
\pagerange{\pageref{firstpage}--\pageref{lastpage}} \pubyear{2008}
\maketitle
\label{firstpage}

\maketitle

\begin{abstract}
Growing Supermassive Black Holes (SMBH) are believed to influence their parent
galaxies in a negative way, terminating their growth by ejecting gas out
before it could turn into stars. Here we present some of the most
sophisticated SMBH feedback simulations to date showing that quasar's effects
on galaxies are not always negative. We find that when the ambient shocked gas
cools rapidly, the shocked gas is compressed into thin cold dense shells,
filaments and clumps. Driving these high density features out is much more
difficult than analytical models predict since dense filaments are resilient
to the feedback. However, in this regime quasars have another way of affecting
the host -- by triggering a massive star formation burst in the cold gas by
over-pressurising it. Under these conditions SMBHs actually accelerate star
formation in the host, having a positive rather than negative effect on their
host galaxies. The relationship between SMBH and galaxies is thus even more
complex and symbiotic than currently believed. We also suggest that the
  instabilities found here may encourage the chaotic AGN feeding mode.
\end{abstract}

\section{Introduction}

Several properties of galaxies correlate with the mass of SMBHs residing in
their centres \citep{Tremaine02,Haering04,CatEtal09}, suggesting a causal
link.  These relations are currently attributed to the appreciable destructive
powers of quasars \citep{SilkRees98,ZK12a}. At high SMBH accretion rates,
quasars launch wide angle outflows with velocity of $v_{\rm out} \sim 0.1 c$
as observed \citep{PoundsEtal03a,PoundsEtal03b,TombesiEtal10} in AGN X-ray
absorption spectra, capable of ejecting all the gas from the host galaxy
\citep{SilkRees98,ZK12a}. Analytical arguments\citep{King03} set the
momentum-driven critical SMBH mass limit, $M_\sigma$, that closely matches the
observed correlations\citep{Tremaine02,Haering04,CatEtal09}. However how the
outflow expels the gas is not clear in detail: previous analytical models
include the essential outflow physics but assume spherical symmetry; numerical
models are 3D but physically over-simplified due to computational constraints
\citep{DiMatteo05,SijackiEtal07,BoothSchaye09}.

As any fast outflow from a point source, the quasar outflow leads to two
shocks, one inner -- the shocked SMBH outflow itself -- and the other outer,
representing the shocked host galaxy gas. The two shocks are separated by a
contact discontinuity \citep[see fig. 1 in][]{ZK12a}.  Previous work has shown
that the radiative cooling of the inner shock has a profound effect on the
ability of the SMBH to drive the gas out of host galaxies. If the radiative
time of the inner shock is short, only the momentum from the primary quasar
outflow is efficient in affecting the host gas, with the mechanical energy
lost to radiation \citep{King03}. When the cooling time is long, the
mechanical energy of the outflow thermalises \citep{King05} and increases the
hot gas pressure inside the quasar-driven bubble, leading to an even more
powerful and rapid $v\sim 1000$ km~s$^{-1}$ outflow \citep{ZK12a}.

Here we present numerical simulations that combine the strengths of the
previous analytical (detailed feedback physics) and numerical (3D geometry)
methods. Our simulations show that the way in which quasar outflows
  affect the gas in the host galaxies also strongly depends on whether the
  {\em outer} shock rapidly cools radiatively or not. In the former case, the
  shocked outer layer is very dense and is unstable to the ``thin shell''
  instabilities previously known from supernova remnant studies
  \citep{Vishniac1983,MacLowEtal89}. These instabilities lead to the shell
  breaking up into thin dense filaments that are resilient to the outward push
  from the quasar outflow but are susceptible to a triggered (or at least
  accelerated) star formation. We propose that the latter process may
  constitute a positive quasar feedback effect on their host galaxies. Below
  we detail where and when this effect may be important.

\section{Model and numerical methods}

Due to computational limitations, sub-grid models are a necessity to describe
SMBH feedback in simulations that span cosmologically interesting
volumes. Depositing energy into the gas nearest to SMBH (e.g., SPH particle
neighbours of the SMBH) is the most widely used method
\citep{DiMatteo05,SijackiEtal07,BoothSchaye09} for such simulations. In this
paper we study AGN feedback on scales of a single galactic bulge, and
therefore we can afford (a) a much higher numerical resolution and (b) a more
detailed physical model for AGN feedback. The outflow physics follows the
\cite{King03,King05,KZP11} model \citep[cf. \S 2.1 in ][]{ZN12a} while the
numerical method follows \cite{ZN12a} exactly (see their \S 3.1), except for
different initial conditions. In brief, hydrodynamics of gas in the host
galaxy and gravity solvers are based on code {\it Gadget}
\citep{Springel05}. In contrast to the nearest neighbour method for AGN
  feedback, we model the propagation of fast ($v_{\rm out} = 0.1 c$) outflow
  from the SMBH explicitly via ``virtual particles'' introduced by
\cite{NayakshinEtal09a}. These particles are emitted isotropically by the
SMBH, and propagate radially outward  with no self-interaction
  allowed. SPH density field is continuously calculated at the particles'
  instantaneous locations, and the interaction with an SPH particle occurs
  only when the latter contains the virtual particle within its SPH smoothing
  kernel. The method has been shown \citep{NayakshinPower10} to reproduce
  analytical results of \cite{King03,King05}, and has been compared with the
  nearest neighbours feedback method by \cite{PowerNK11}. \cite{ZN12a} further
  included optically thin photo-ionisation and inverse Compton heating of the
  galaxy's gas by the quasar radiation field \citep{SazonovEtal05}; this
  heating may be an important feedback effect in gas-poor epochs
  \citep[e.g.,][]{CiottiOstriker07a}.

Additionally, not only the momentum but also a fraction $g_{\rm E} = \exp
[- R_{\rm ic}/R]$ of the virtual particles' mechanical energy is passed to the
  ambient SPH particles, where $R_{\rm ic}\approx 0.5$~kpc (for the parameters
  chosen below) is the inner shock cooling radius of the inner AGN outflow
  shock \citep{King03,King05}. This radius is not to be confused with the
  outer shock cooling radius, $R_{\rm oc}$, introduced later in the Discussion
  section of the paper. Note that $g_{\rm E}(R\ll R_{\rm ic})\rightarrow 0$,
  while $g_{\rm E}(R\gg R_{\rm ic})\rightarrow 1$. This energy deposition
  prescription mimics the transition from the radiatively efficient inner
  shock regime (small radii), when only the momentum of the SMBH outflow is
  passed to the gas, and the radiatively inefficient regime (large radii),
  when both energy and momentum are available to drive the host's gas out
  \citep[see][]{KZP11}. \cite{FQ12a} have recently noted that if ions decouple
  thermally from electrons in the inner shock then the radiative cooling
  becomes far less efficient than obtained with a one-temperature model for
  the shock. The inner cooling radius $R_{\rm ic}$ in this case is negligibly
  small. We note that this (if correct) would introduce quantitative but not
  qualitative changes to our main results below, as we experimented with
  shells initially much smaller and also much larger than $R_{\rm oc}$
  (to be reported elsewhere).

To demonstrate our points as clearly as possible, the host galaxy is modelled
here by a fixed singular isothermal sphere potential softened at the centre
where the SMBH is located. The ambient gas is initialised as a sphere with
inner radius $R_{\rm in}$ and outer radius $R_{\rm out} = 10 R_{\rm in}$, at
rest, and with the density profile given by 
\begin{equation}
\rho_g\left(R\right) = f_g \rho_0\left(R\right) = {f_g
\sigma^2 \over 2\pi G R^2} \;, 
\label{rhog}
\end{equation}
with $\rho_g(R) = 0$ for $R < R_{\rm in}$ and $R > R_{\rm out}$; here $\sigma
= 141$~km~s$^{-1}$ is the velocity dispersion of the potential, $f_g = 0.16 f$
is the gas ratio of the gas density to the underlying density of stars and
dark matter; $f>0$ is a free parameter. Initial gas temperature is set such
that the sound speed is equal to $\sigma$. At time $t=0$, the quasar outflow
is turned on. The critical $M_\sigma$ mass above which the SMBH outflow should
expel the gas for this potential \citep{NayakshinPower10} is $M_\sigma \approx
1.5\times 10^8 f \msun$.

We fix the quasar momentum outflow rate and the SMBH mass in order to avoid
model-dependent complications near SMBH. To avoid increasingly short
time-steps at small radii, we use the accretion radius approach
\citep{PowerNK11} and remove SPH particles from the simulation that are closer
than $30$ pc from the SMBH. The densest and coldest self-gravitationally bound
regions would continue to contract to arbitrarily large densities, a process
which physically terminates in star formation. When gas density exceeds 200
times the potential's tidal density $\rho_0$, and the Jeans mass in the clump
is smaller than the SPH kernel (40 particle neighbours) mass, SPH particles
are turned into collisionless star particles. Our treatment of star formation
neglects feedback effects from massive stars; this does not alter our main
conclusions.

\section{Results for spherical shells}\label{sec:results}

We first present two simulations that exemplify the effects of the quasar
outflow on the host galaxy's gas in two extremes, e.g., (i) when the shocked
gas can cool rapidly and (ii) when radiative cooling is inefficient (low gas
density).

\subsection{Thin shell instabilities}

\begin{figure*}
\centerline{\psfig{file=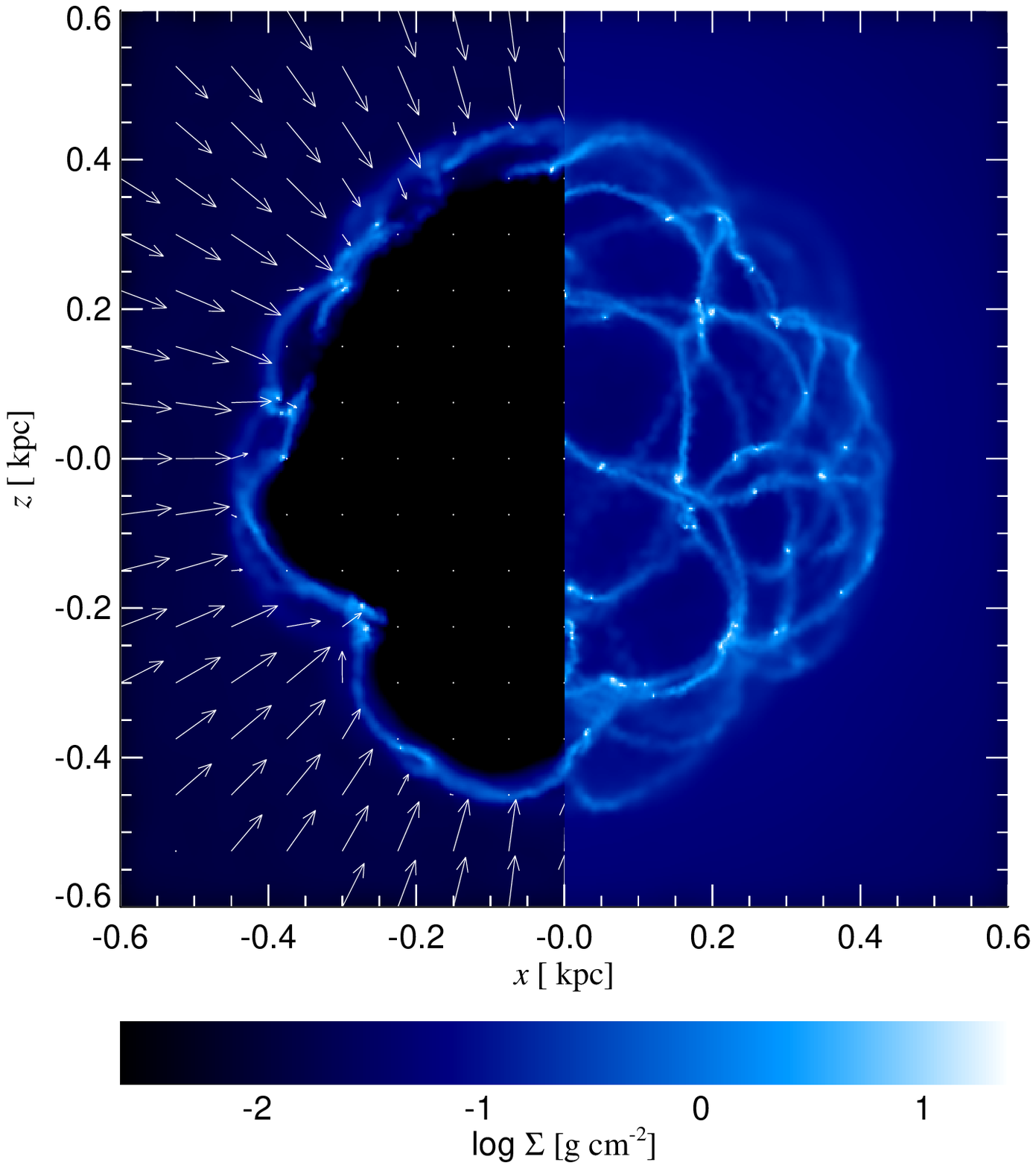,width=0.45\textwidth,angle=0}
\psfig{file=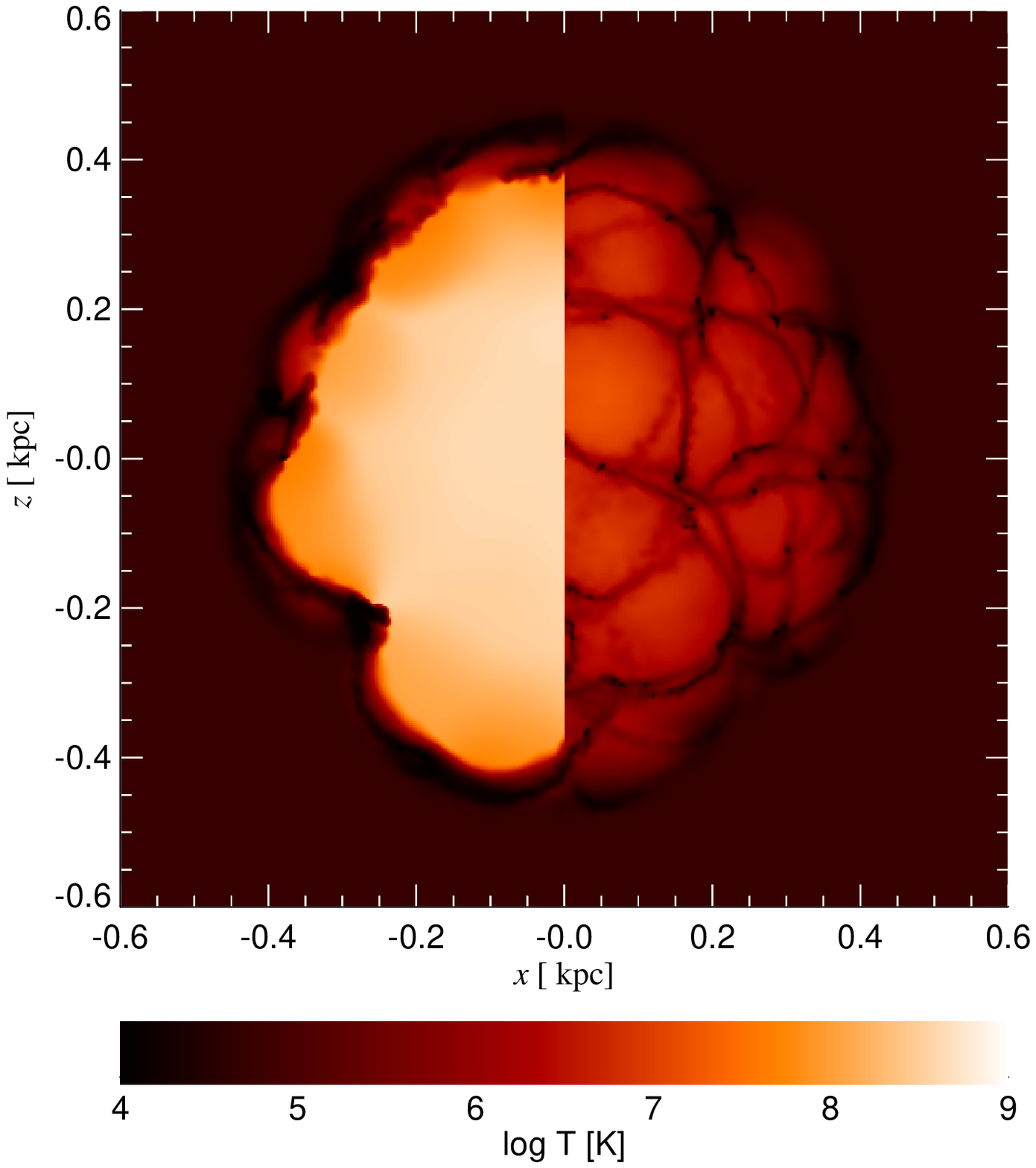,width=0.45\textwidth,angle=0}}
\vskip -1.4 cm
\centerline{\psfig{file=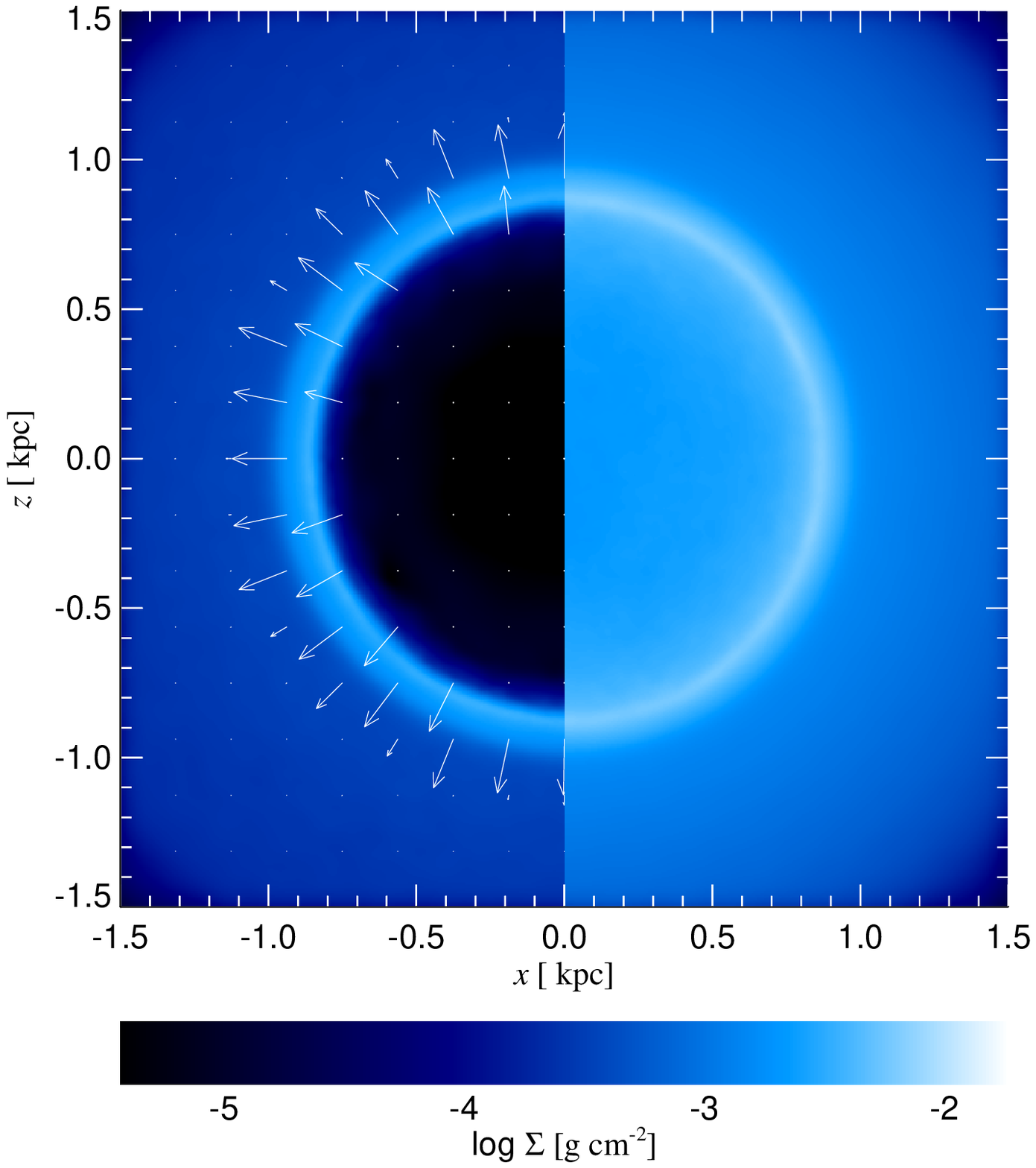,width=0.45\textwidth,angle=0}
\psfig{file=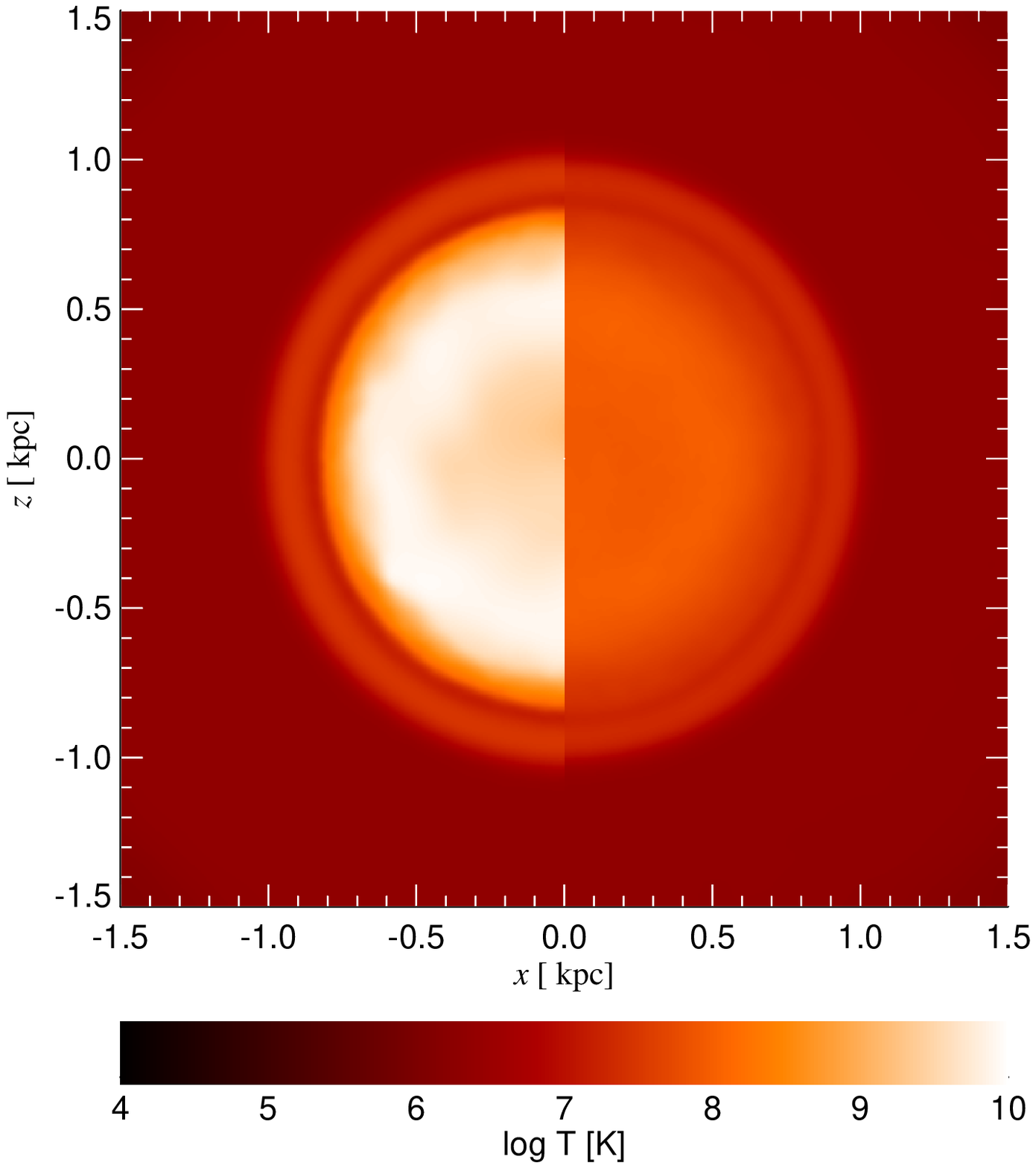,width=0.45\textwidth,angle=0}}
\vskip -1.4 cm
\caption{Gas outflows for a gas-rich simulation F1 (top panels) at time $t=
  3.8$~Myr (top) and same for a gas-poor simulation F0.03 at $t=0.57$~Myr
  (bottom panels). The left panels show projected gas density and the right
  ones show gas temperature. To expose the 3D nature of the flows, the left
  sides of each panel show geometrically thin ``wedge-slice'' projections of
  the gas flow, where only regions $y^2/(x^2 + y^2 + z^2) < 1/16$ are shown;
  the right-hand sides of the panels show full cube projections.  The shell in
  the gas-rich simulation cools rapidly and becomes unstable to thin shell
  instabilities \citep{Vishniac1983}. The shell in the gas-poor simulation cools
  very slowly, and is thus hot and radially extended. This shell is stable and
  remains spherically symmetric as the gas is driven away.}
\label{fig:r2f1_f0.03}
\end{figure*}

Figure \ref{fig:r2f1_f0.03} shows the density (left column) and temperature
(right column) of the ambient gas shocked by the outflow in simulation F1 in
the gas-rich epoch ($f = 1$, top panels), and in simulation F0.03,
corresponding to a gas-poor epoch regime ($f = 0.03$, bottom panels). For both
simulations, the SMBH mass is $\mbh = 10^8\msun$ and the outer shell radius is
$R_{\rm out} = 2$~kpc. The host galaxy is rapidly emptied of gas in F0.03; the
shock front velocity is about 1500 km s$^{-1}$, so that the shock expanded to
$R\sim 1$ kpc by the time of the snapshot, $t=0.57$~Myr. In contrast to that,
the same SMBH outflow in the gas-rich simulation F1 stalls: the radius of the
shock front is $R\sim 0.4$ kpc at time $t=3.8$~Myr.  This immense difference
is due to two factors, both significant here: (i) the weight of the gas larger
by a factor of $\approx 30$ in F1 than in F0.03, and (ii) the inner shock
  remains in the momentum driven regime in F1, since the shell stalls at $R <
  R_{\rm oc}$, whereas the shock becomes non-radiative and thus energy driven
  in simulation $F0.03$. These results are consistent with the analytical
expectations for spherically symmetric shells \citep{ZK12a}.

An unexpected result is that while the shocked shell is quasi-spherically
symmetric in F0.03, it goes violently unstable in F1. In hindsight, the result
is completely analogous to the stability of supernova remnant shells. {\em
  Cold} dense swept-up shells are strongly unstable to Rayleigh-Taylor and
\cite{Vishniac1983} instabilities \citep{MacLowEtal89}, whereas {\em
  non-radiative hot and geometrically thick} shocks are stable (compare fig. 3
in \cite{MacLowEtal89} with the rest of their figures). Physically, radiative
cooling is key for development of the instabilities because any initial
density inhomogeneities are amplified when denser regions cool and get
compressed by the high pressure surroundings; without cooling these density
inhomogeneities tend to diffuse away.

The cold dense filaments are crucial to understanding of SMBH feedback on host
galaxies and also the SMBH feeding, as we shall see.

\subsection{Accelerated star formation}

\begin{figure*}
\centerline{\psfig{file=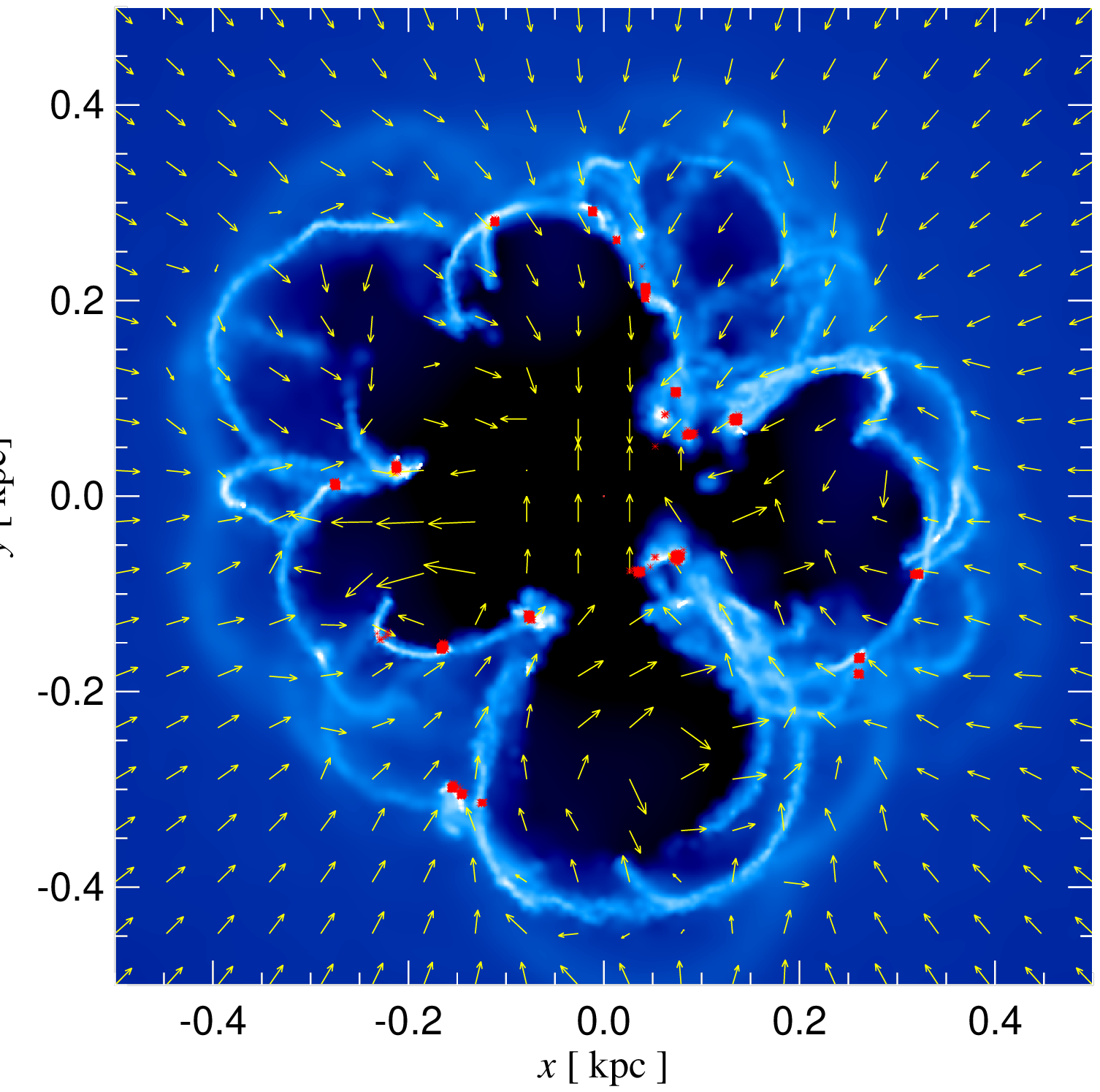,width=0.49\textwidth,angle=0}
\psfig{file=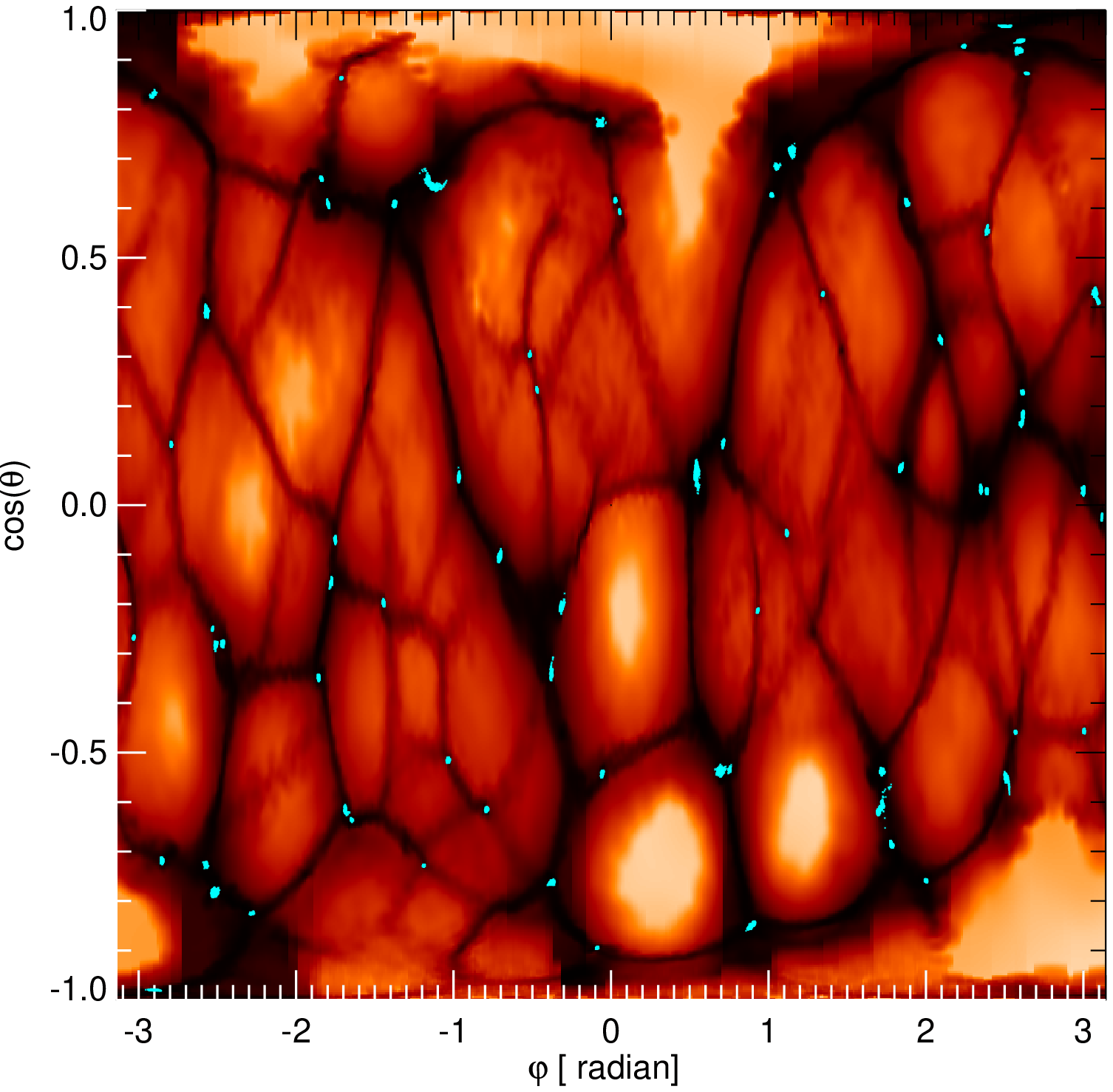,width=0.49\textwidth,angle=0}}
\caption{Left panel: ``Wedge projection'' (see fig. 1) of gas density and
  velocity vectors for simulation F1 at time $t=5.2$~Myr. The high density
  regions of the shell receive less quasar feedback per unit mass and thus
  fall inward. Red symbols show newly born stars that form within the dense
  filaments. Most of the stars form in clusters which are disrupted when
  passing through the central regions of the galaxy. Right panel: Temperature
  of the gas as seen from the SMBH, averaged along rays, for a shell $0.2 < r
  < 0.4$ kpc, at time $t=4.3$~Myr, and as a function of the two spherical
  angles in spherical coordinates, defined in the usual way: $\cos\theta =
  z/\sqrt{x^2 + y^2 + z^2}$, and $\sin\phi = y/\sqrt{x^2 + y^2} $. Cyan
  symbols show locations of the stars. These form exclusively inside the
  density peaks (corresponding to minima in temperature), which occur most
  frequently at the intersections of the filaments. The colour scales for the
  panels are same as they are in figure 1 for simulation F1.}
\label{fig:globe}
\end{figure*}

The cold dense filaments in simulation F1 are the birth place for newly born
stars.  The right panel of Figure \ref{fig:globe} presents the gas temperature
projected along rays as seen from the SMBH for simulation F1. The shell is a
web of cold dense filaments with the strongest density peaks occurring at
filament intersections. The left panel of the figure shows the density and
velocity field at the time when dense filaments are starting to fall
inward. Stars (red or cyan dots) are born inside the densest and coldest
filaments.

Physically, quasar shocks are able to trigger star formation, or accelerate
it, by compressing the cold gas in the hosts because the pressure produced by
the quasar outflow is far higher than the maximum gas pressure inside the
pre-quasar host, $P_{\rm max} \sim f_g \rho_0 \sigma^2$ (if the gas were
hotter, $c_s^2 > \sigma^2$, a thermal pressure-driven wind would result). The
pressure in the shocked ambient gas and inside the hot bubble inflated by the
quasar is $P_{\rm sh} \sim f_g \rho_0 v_e^2$, where $v_e \gg \sigma $ is the
energy-driven shell expansion velocity calculated in \cite{KZP11}. In the
energy-driven regime, the quasar's bubble pressure is higher than $P_{\rm
  max}$ by the ratio
\begin{equation}
{P_{\rm sh} \over P_{d}} = \left({ v_e \over \sigma}\right )^2 \sim 24\;
\sigma_{150}^{-2/3} f^{-2/3}\gg 1\;.
\label{pratio}
\end{equation}
To exemplify the quasar's ability to trigger star formation further, consider
the effects of the quasar outflow on a starburst gas disc. Such discs may have
star formation rates as high as $\sim 10^3 \msun$~yr$^{-1}$ while
self-limiting their star formation rates \citep{Thompson05} by star formation
feedback. Their midplane pressure is $P_d \sim \rho_0 \sigma^2 f_d^2$, where
$f_d < f_g$ is the disc mass fraction at a given $R$. This is smaller than
$P_{\rm max}$, so we see that even the starburst-supported discs are
compressed to higher densities by quasar outflows. Since star formation rates
are proportional to $P_d$ in these discs \citep{Thompson05}, disc compression
due to quasar shock results in an even faster starburst in the disc.

\subsection{Chaos and order in SMBH accretion}

SMBH accretion is still far from understood. One of the significant challenges
is the ``star formation catastrophe'' of self-gravitating parsec scale discs
\citep{Paczynski78,NayakshinEtal07} in which gas is consumed more rapidly in
star formation than it is fed to the SMBH. One proposed solution is disc
self-regulation \citep{Goodman03}, and the other is ``chaotic accretion'', in
which angular momentum of gas is cancelled in shocks rather than transferred
away \citep{KingPringle06}. Chaotic SMBH feeding is suggested to encourage a
more rapid SMBH growth in the early Universe \citep{KingPringle06} by limiting
SMBH spin and may also explain \citep{NPK12a} why SMBH do {\em not} correlate
well with properties of galactic discs \citep{KormendyEtal11}. Such flows
could develop as a result of messy initial conditions or supernova-driven
turbulence in the bulge \citep{HobbsEtal11}.

Thin shell instabilities \citep{Vishniac1983}, re-discovered here in the SMBH
growth context, are thus an additional way of developing such multi-phase
chaotic inflows {\em starting from non-turbulent spherically symmetric}
initial conditions. The resulting gas flows are indeed highly variable as
infalling fingers (see figure 2) shadow the ambient gas in an unpredictable
way. On the other hand, if the ambient gas possesses symmetry before SMBH
feedback is turned on, then there may be preferred inflow directions and an
orderly flow/outflow develops in this case: for example, see simulations by
\cite{ZN12a} for the {\it Fermi Bubbles} in the centre of the Milky Way.
Further simulations with a range of realistic initial conditions are needed to
investigate the importance of chaotic inflows for SMBH feeding.

\section{Implication for non-spherical geometries}\label{sec:implications}

Having investigated an initially spherically symmetric shell in \S
\ref{sec:results}, we now wish to ask what happens in a more realistic
case. Based on the earlier results, we expect that deviations from spherical
symmetry in initial conditions will be amplified by the thin shell
instabilities when the ambient shocked gas is able to cool rapidly. As before,
the flow may break into dense inflowing and lower density outflowing regions
or directions. If angular momentum is important then the denser flows may
settle into a circularised disc.

\begin{figure*}
\centerline{\psfig{file=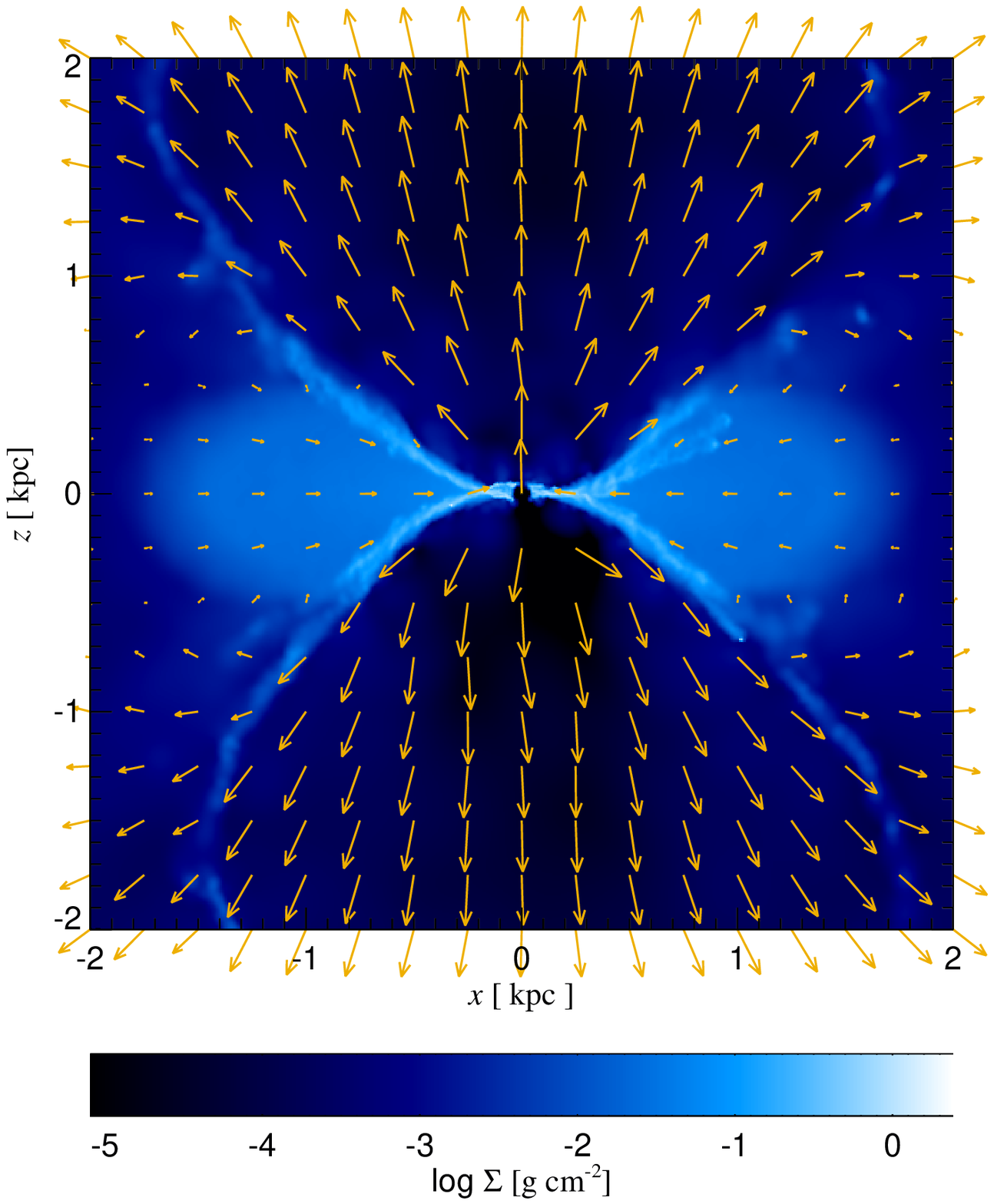,width=0.34\textwidth,angle=0}
\psfig{file=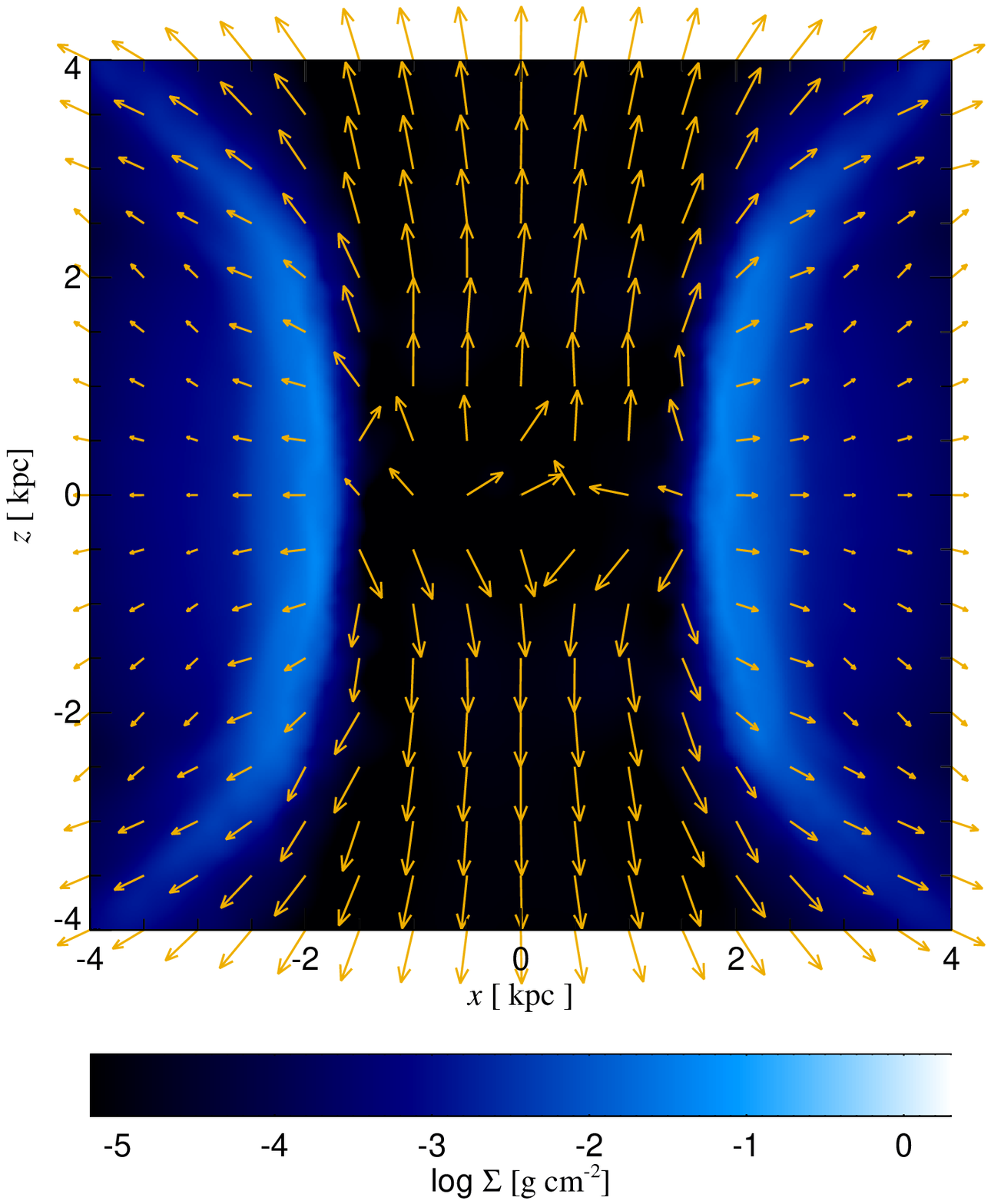,width=0.34\textwidth,angle=0}
\psfig{file=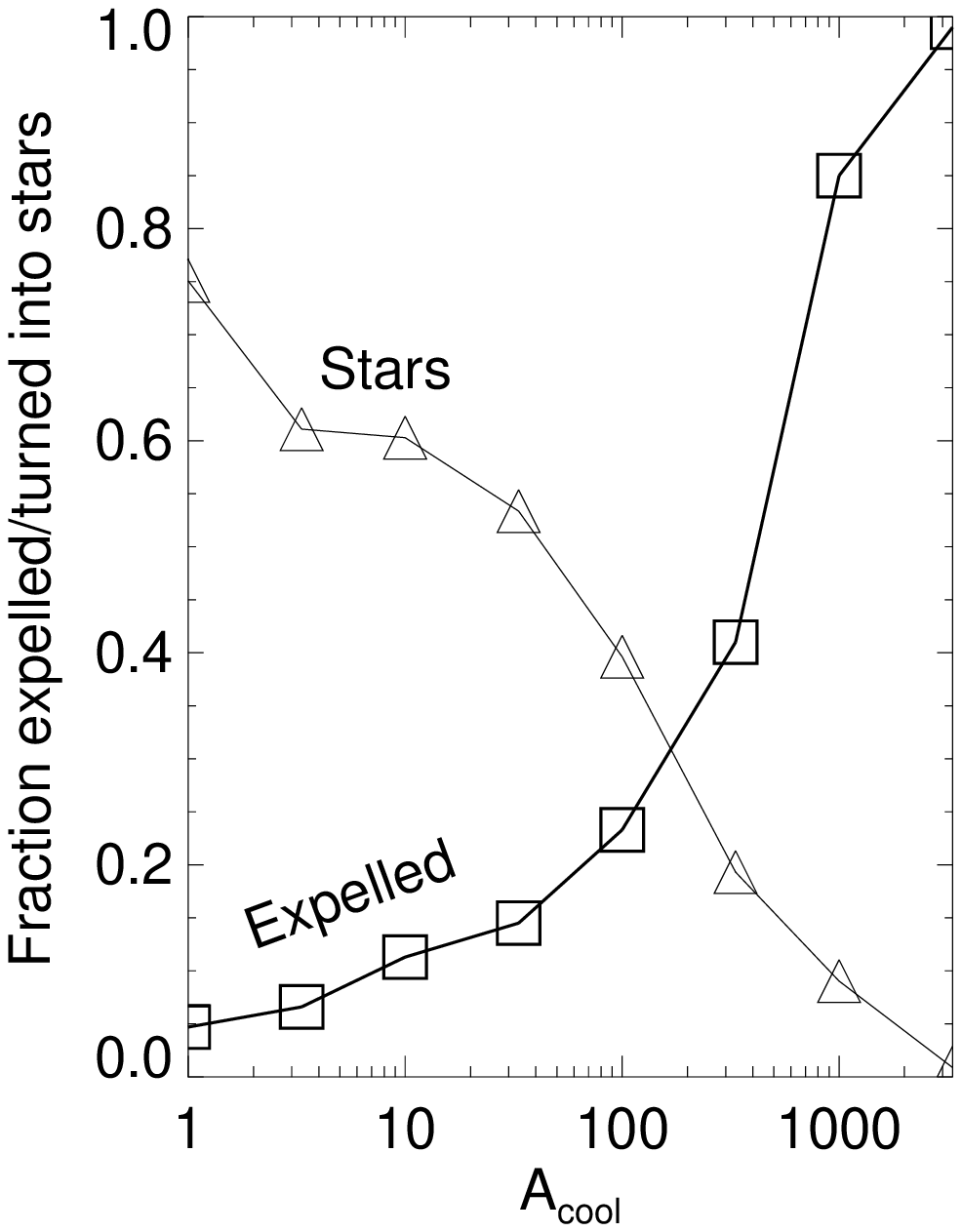,width=0.32\textwidth,angle=0}}
\caption{Quasar feedback on an initially elliptically-shaped and slowly
  rotating shell in the cases with the proper radiative cooling on for the
  shocked gas (left panel) and for the case when radiative cooling is switched
  off (middle panel). In both cases gas is expelled along the vertical
  direction. However, with cooling on, the gas settles into a high density
  disc that is resilient to quasar feedback. Very little gas is ejected from
  the galaxy in this case; most is turned into stars or accreted through the
  inner boundary condition. In contrast, with cooling off, all of the ambient
  gas is expelled to infinity because no high density features form; hot and
  diffuse gas is much more susceptible to quasar feedback. The right panel
  shows how the fraction of the shell expelled to infinity and turned to stars
  at the end of the simulations ($t=9$~Myr) change as radiative cooling is
  progressively suppressed by factor $A_{\rm cool}$ (the left panel
  corresponds to $A_{\rm cool} = 1$, and the middle panel to $A_{\rm cool} =
  \infty$). More and more gas is ejected rather than turned into stars as
  cooling is suppressed.}
\label{fig:ellipse}
\end{figure*}

We wish to emphasise the fact that the cooling rate of the shocked ambient gas
is key to this behaviour. Figure \ref{fig:ellipse} presents two contrasting
simulations with same initial conditions but with ambient gas radiative
cooling either on (left panel) or off (middle panel). The initial conditions
are obtained from the spherical shell used in F1 by a linear transformation $z
\rightarrow z/2$ which obviously makes an elliptical rather than a spherical
initial shell.  The shell also slowly rotates, $v_{\rm rot} = 0.3 v_{\rm
  circ}$, where $v_{\rm circ} = 2^{1/2}\sigma$ is the circular velocity for
the potential, with the angular momentum vector pointing in the positive $z$
direction. Finally, the SMBH mass is slightly increased to $\mbh = 1.5\times
10^8\msun$, the $M_\sigma$ mass for this potential.

The left panel of Figure \ref{fig:ellipse} shows that the outflow is
channelled along the symmetry axis. However, as the ambient gas cools, most of
it collapses onto the symmetry plane, becoming resilient to feedback and
instead mainly turning into stars. This is in a stark contrast to a simulation
shown in the middle panel of the same figure, where the radiative cooling in
the ambient gas is artificially turned off. All of the gas is expelled in this
case because the shocked ambient gas remains {\em hot} and always presents a
large enough surface area for the SMBH feedback to work on.

To quantify this cooling-regulated behaviour, we introduced a free parameter
$A_{\rm cool} \ge 1$, which is used to artificially reduce the radiative
cooling rate function $\Lambda \rightarrow \Lambda/A_{\rm cool}$ in a set of
numerical experiments which share same initial conditions as explained
above. The left panel of Figure \ref{fig:ellipse} corresponds to $A_{\rm
  cool}=1$, e.g., the proper unmodified cooling rate, whereas the middle panel
of the figure corresponds to $A_{\rm cool} = \infty$ (e.g., cooling in the
ambient gas is turned off).  The right panel of figure \ref{fig:ellipse} shows
the fractions of gas expelled and turned into stars as functions of $A_{\rm
  cool}$. We see that quasar outflows mainly turn gas into stars when the
shocked gas cools rapidly but is able to expel most or even all of the gas if
the shocked gas is unable to cool (large $A_{\rm cool}$).

\section{Discussion}

 We presensted simulations of the interaction between the fast outflow
 emanating from a quasar and the ambient gas in the host galaxy utilising the
 he virtual particle method of \cite{NayakshinEtal09a}. We believe the method
 is an improvement \citep[see also][]{PowerNK11} on the sub-grid
 nearest-particle methods since the SMBH outflow is explicitly modelled to
 track where and how it impacts the ambient gas, and whether the outflow's
 momentum, energy, or both are passed to the ambient gas. In particular,
 consider the left panel of figure \ref{fig:globe}. The cold filaments subtend
 a small solid angle as seen from the quasar location and should thus
 encounter a small fraction of the quasar outflow only, with most quasar
 feedback impacting on the low density gas instead. This is correctly captured
 by our virtual particle method. In contrast, the nearest neighbour methods
 may spread {\em all of the quasar feedback energy} to the cold filaments
 because they tend to be closer to the SMBH.

Our key new result is the finding that the response of the ambient gas to the
fast outflow of gas from the quasar depends critically on whether the ambient
gas is able to cool rapidly or not. Generally, in the latter case the
analytical criteria, such as the outflow momentum (thrust) balanced against
the weight of the gas in the bulge are sufficient to establish whether the
shell is driven outward, stalls or collapses
\citep{King03,King05,KZP11}. However, when the outer shock is radiative, even
initially spherically symmetric shells develop a multi-phase structure. Denser
regions cannot be driven outward easily. Quasar outflows affect these high
density regions most effectively not by expelling them outward but by
compressing them and triggering (or at least accelerating) star formation
there.

Let us now quantify the conditions when the ambient shocked gas is able to
cool rapidly. Bremsstrahlung is the most significant cooling function for the
problem at hand, with $\Lambda(T) = \Lambda_0 T^{1/2}$, $\Lambda_0 = 3.8
\times 10^{-27}$ erg~cm$^3$~s$^{-1}$. The shocked gas cooling time, $t_{\rm
  cool} = 3 kT/n\Lambda(T)$, where $n$ and $T$ is the shocked gas electron
density and temperature, respectively. In the strong shock limit, $n = 4
  \rho_g(R)/(\mu m_p)$, where $\mu = 0.63$ is the mean molecular weight and
  $\rho_g(R)$ is the pre-shock gas density given by equation \ref{rhog}, and
  $T = (3/16) \mu m_p v_e^2/k_B$.  To determine whether the quasar shock is
radiative or adiabatic, $t_{\rm cool}$ should be compared with the shocked
shell flow time scale, $R/\dot R$, where $\dot R = (4/3)v_e$ is the outer
shock velocity \citep{ZK12a}:
\begin{equation}
{t_{\rm cool}\over R/\dot R} = 0.83 \; { 3.8\times 10^{-27} \over \Lambda_0}
\; f^{-5/3} \sigma_{150}^{-2/3} \;\left({
  \mbh \over M_\sigma}\right)^{2/3}\; R_{\rm kpc}\;.
\label{tratio}
\end{equation}
 At a fixed fraction $f$, the ``outer cooling radius'', $R_{\rm oc}$, the
 radius which separates the rapidly cooling outer shock regime from the slowly
 cooling one, can be defined by requiring that the ratio in equation
 \ref{tratio} be equal to one. The result is:
\begin{equation}
R_{\rm oc} = 1.2 \;\hbox{kpc}\; f^{5/3} \sigma_{150}^{2/3} \;\left({
  M_\sigma\over \mbh}\right)^{2/3}\; {\Lambda_0 \over 3.8\times 10^{-27} }\;.
\label{roc}
\end{equation}
In the early gas rich epochs (which we operationally define as
$f > 1$) SMBHs affect their hosts both positively via accelerating star bursts
at ``small'' radii (at $R \simlt R_{\rm oc}$), and negatively at larger radii
where the gas is more likely to be blown out. Interestingly, the powerful
starbursts in submillimeter galaxies (e.g., galaxies forming stars at rates as
high as $\sim 10^3 \msun$~yr$^{-1}$ at redshift $z \sim 2$) have been resolved
by \cite{TacconiEtal06a} to have intrinsic sizes of a few kpc, comparable to
$R_{\rm oc}$ in the gas-rich epoch.

In the later gas-poor epochs, when $f\ll 1$ everywhere, and $\mbh \simgt
M_\sigma$, it follows from equation \ref{roc} that $R_{\rm oc}\ll 1 $~kpc, so
that shocked gas is unlikely to clump up into high density features over most
of the observationally interesting scales in the host galaxy. In this regime
quasars should be better able to sweep the hosts clear of gas if they are
activated for long enough. This is the widely appreciated negative quasar/AGN
feedback mode, and it probably explains why the present day ellipticals are
``red and dead'' and why their SMBHs are not very bright
\citep{SchawiskiEtal09}.

\section{Conclusions}

Our results broaden the spectrum of possible effects that SMBHs have on their
host galaxies, showing that it is not all about ejecting the gas out. We
  found that it is increasingly difficult to drive gas out of the hosts at
  small radii, $R < R_{\rm oc}$ (equation \ref{roc}), in the early gas-rich
  epoch. Even in an initial spherical symmetry the thin and dense outer shell
  breaks into filaments and clumps that are hard to drive away but which can
  be readily over-pressurised by the hot quasar wind and which result in
  enhanced star formation rates. The latter process is a form of positive
  rather than negative AGN feedback on host galaxies \citep[we note that
    similar statements were already made by][based on their analytical
    arguments]{SilkNorman09}. Non-spherical initial shell geometries only
  strengthen these conclusions as the multi-phase inflow-outflow separation
  becomes even more robust (see \S \ref{sec:implications}).

We also found that in later gas-poor epochs, when the shocked host's gas
cannot cool rapidly (cf. Discussion), the inhomogeneities do not develop
(cf. the bottom panel of fig. 1) or do not grow for non-spherical initial
conditions (cf. figure 3, the middle panel). The SMBH is then an efficient
negative feedback engine as has been appreciated for a long time now.

\del{Finally, it is instructive to compare these results with what is known for the
other and better studied sources of negative feedback in galaxy formation --
massive stars. These are found \citep{DeharvengEtal05} to produce a {\em
  positive feedback also:} shocks driven by supernova explosions, stellar
winds or photo-ionisation into the ambient gas pressurise the latter to high
densities and can result in a triggered star formation. SMBHs and massive stars
are thus similar in this regard, changing the nature of their impact on the
surrounding gas depending on the physical conditions in the latter.}

\section{Acknowledgments}

Andrew King is thanked for helpful comments and encouragement.  Theoretical
astrophysics research at the University of Leicester is supported by a STFC
Rolling grant.  This research used the ALICE High Performance Computing
Facility at the University of Leicester.  Some resources on ALICE form part of
the DiRAC Facility jointly funded by STFC and the Large Facilities Capital
Fund of BIS.

\label{lastpage}

\end{document}